# Strain and Crystallographic Identification of the Helically Concaved Surfaces of Nanoparticles

Sungwook Choi[1]†, Sang Won Im[2]†, Ji-Hyeok Huh[3], Sungwon Kim[1], Jaeseung Kim[1], Yae-Chan Lim[2], Ryeong Myeong Kim[2], Jeong Hyun Han[2], Hyeohn Kim[2], Michael Sprung[4], Su Yong Lee[5], Wonsuk Cha[6], Ross Harder[6], Seungwoo Lee[3,7], Ki Tae Nam[2]*, Hyunjung Kim[1]*

[1]Department of Physics, Sogang University, Seoul 04107, Korea.
[2]Department of Materials Science and Engineering, Seoul National University, Seoul 08826, Korea
[3]KU-KIST Graduate School of Converging Science & Technology, Korea University, Seoul 02481, Korea
[4]Deutsches Elektronen-Synchrotron (DESY), Hamburg 22607, Germany
[5]Pohang Accelerator Laboratory, Pohang 37673, Korea
[6]Advanced Photon Source, Argonne National Laboratory, Argonne, Illinois 60439, USA
[7]Department of Integrative Energy Engineering and KU Photonics Center, Korea University, Seoul 02481, Korea
†These authors contributed equally to this work
*Corresponding author. Email: nkitae@snu.ac.kr, hkim@sogang.ac.kr

Identifying the three-dimensional (3D) crystal-plane and strain-field distributions of nanocrystals is essential for optical, catalytic, and electronic applications. Here, we developed a methodology for visualizing the 3D information of chiral gold nanoparticles with concave gap structures by Bragg coherent X-ray diffraction imaging. The distribution of the high-Miller-index planes constituting the concave chiral gap was precisely determined. The highly strained region adjacent to the chiral gaps was resolved, which was correlated to the 432-symmetric morphology of the nanoparticles and its corresponding plasmonic properties were numerically predicted from the atomically defined structures. This approach can serve as a general characterization platform for visualizing the 3D crystallographic and strain distributions of nanoparticles, especially for applications where structural complexity and local heterogeneity are major determinants, as exemplified in plasmonics.

**One-Sentence Summary:** Crystallographic indexing and strain analysis of a chiral nanoparticle developed morphology-plasmonic property correlation.

The three-dimensional (3D) distribution of the exposed surfaces on a nanoparticle (NP) is a determinant of its catalytic, optical, and electronic properties (*1–5*). In this regard, the exact indexing of crystallographic-planes and determining the corresponding strain distribution inside or on a single NP are important for correlating the properties with the overall morphology and further tuning the stabilized planes by surface doping or ligand coordination (*6–10*). However, available methods are limited for revealing the three-dimensional, local arrangement of nanometric features, especially for hidden and concave surfaces. An important but unresolved example is plasmonic NPs, whose light-matter interactions at the subwavelength scale can be tailored by controlling the facets, vertices, and morphology of the metallic NPs. According to recent studies, even an atomic-scale topological protrusion of plasmonic NPs can play a pivotal role in squeezing and focusing a photon into a pico-volumetric space (referred to as pico-cavity) (*11, 12*). This otherwise impossible light-matter interaction can be further extended to plasmonic chiro-optical properties (e.g., optical chirality) (*13, 14*), which has implications for novel applications, such as metamaterials and chiral sensing. An engraving of an atomic-scale 3D chiral morphology on the surface of plasmonic NPs serves as a vista for achieving unnatural optical chirality in the visible regime (i.e., optical helicity density, that is, the projection of the spin angular momentum density on the momentum direction) (*1, 15*).

Although tomographic imaging with electrons (*16, 17*) can be used to obtain atomic resolution, it has limited inspection depth and a trade-off between the resolution and object size. These are the main obstacles in the imaging and crystallographic identification of NPs larger than a few tens of nanometers. In particular, for NPs with complex 3D geometries, the loss of 3D information owing to the limited tilting angle of the sample, which is often referred to as the "missing wedge" (*18*), makes it challenging to obtain accurate imaging results and subsequently use the NPs for quantitative plasmonics. In this regard, Bragg coherent X-ray diffraction (CXD) imaging (BCDI) (*19, 20*) could be a generalized platform for the 3D imaging of NPs with deep concave morphologies and hidden surfaces. Owing to the considerable penetration depth of X-rays and the small angular rotation of the samples required to measure the entire 3D CXD pattern, BCDI is a suitable technique for imaging the complex morphologies of NPs from 100 nm to a few micrometers in size. More importantly, the nature of Bragg diffraction allows BCDI to be used to identify crystallographic information (*21*). The defined scattering geometry of the image provides the reference crystallographic orientation to determine the surface Miller indices and the reconstructed phase in BCDI provides lattice deformation with picometer-scale sensitivity.

As an effort to develop the general method for analyzing complex 3D NPs, we investigated the crystallographic identification of the entire surface and strain-field analysis of a chiral gold (Au) NP, whose properties are exotic and have been considered difficult to characterize in terms of atomic surface structures. The investigated NP, which is called 432 helicoid III was a single-crystal with the features of uniformly formed 3D concave chiral gaps with complex geometry and surface Miller indices (*1*). We adopted 432 helicoid III as a model because its most complex morphology and chiral plasmonic behavior is not fully understood at the single NP level. The chiral gaps result from the interaction between chiral peptide molecules and the concave high-Miller-index planes exposed during the growth of the NP (*22, 23*). Although it is of interest to correlate its structure and functional properties, the exact gap structure and the distribution of the surface Miller indices remain unclear, limiting an in-depth understanding of the mechanisms of chiral morphology formation. We found the following important characteristics of the chiral Au NP: (i) the 3D gap surfaces consisted of a complex composition of high-Miller-index planes with a mixed distribution of R and S chirality, (ii) the overall Miller-index distribution of the gap surfaces possessed 432-symmetry identical to the NP morphology, and (iii) the strain-field inside the NP was also symmetric, indicating a correlation between the lattice deformation and concave morphology.

We investigated the CXD patterns of 432 helicoid III and then reconstructed them into 3D images. Coherent X-rays were focused on NPs, loaded with the (100) facets facing upward on the substrate (Figs. 1A and S1). The 3D CXD pattern was collected by successive measurements of slices through reciprocal space in the vicinity of the Au (200) Bragg peak by rocking the Bragg angle of the sample. To satisfy the (200) Bragg reflection geometry, a focused X-ray was incident at 19.8° with an X-ray energy of 8.84 keV. Under this condition, the wavevector transfer (**Q**) is parallel to the lattice vectors perpendicular to the (100) facets pointing upward and the arbitrary rotation of the NP around **Q** is allowed. The measured Au NPs (432 helicoid III) exhibited chiral gaps with concave surfaces positioned at the edges of the cubic morphology, as shown in the scanning electron microscopy (SEM) image in Fig. 1B. Owing to its complex morphology and the directions of the surface facets, 432 helicoid III showed a 3D CXD pattern with complex interference fringes in multiple directions (Fig. 1C). These directions were mainly distributed in the high-Miller-index region on stereographic projections (*24*), in contrast to those of cubic NPs (Figs. 1D and S2)

From the 3D CXD pattern, the Bragg electron density of the NP was reconstructed through an iterative phase-retrieval process (*25*). By taking the absolute value of the complex electron density, the 3D morphology of 432 helicoid III in Fig. 2A, which is consistent with the SEM image, shows the spatial distribution of the Miller indices resolving the complex geometry. We determined the crystallographic orientation of the NP from the **Q** vector and the surface normal direction (**N**) parallel to the lattice vector, which was determined from fitting based on the Terrace-Step-Kink model (*26*), which maximizes the terrace ratio of the surface (Fig. S3). The K means-based algorithm was used to avoid artifacts from the pixilation error and limited spatial resolution (Fig. S4). Two types of regions of the 432 helicoid III surface, flat outer facets and concave chiral gaps, were clearly distinguished based on the crystallographic features. The flat exterior facets mostly consisted of {100}, as expected from the cubic morphology. In contrast, high-Miller-index facets were observed on the surface of the concave gaps, in addition to low-Miller-index facets such as {110} and {111}. Because the gap has tilted and highly curved surfaces, high-Miller-index surfaces with intermediate slopes are necessary for the continuous surface. The fourfold, threefold, and twofold rotation symmetry axes of 432 helicoid III were also identified, corresponding to [100], [111], and [110], respectively (Fig. S5). The chiral gaps were positioned at the [110] edges of the cubic morphology whose surfaces were facing [100]. Each gap had a concave geometry carved toward the inside and tilted inward, determining the overall symmetry of the NP. Each gap can be considered to be a twofold symmetric motif that is located at the center of 12 edges by fourfold and threefold symmetry, combining to form 432-symmetry. Although it is not ideally symmetric, it is in contrast to the fact that the geometry of a typical single-crystalline NP follows its lattice symmetry, e.g., a cube NP has the same $4/m\bar{3}2/m$ symmetry as the face-centered cubic crystal lattice. This difference originates from the effect of the chiral peptide, which breaks the mirror symmetry of the crystal facets. The kink density was calculated from the extracted Miller-index and mapped on the 432 helicoid III morphology (Fig. 2B). It was confirmed that chiral kink sites were mainly exposed on the surfaces of the concave gaps. From this analysis, we can obtain important evidence for understanding the synthetic mechanism during growth. The lesson is that the chiral gap structure comprises symmetric-preserved high-Miller-index surfaces.

By converting the surface normal vectors into stereographic projections, the distributions of the surface Miller indices are shown in Fig. 2C. The surface area of each Miller-index with color-scale is shown for the overall distribution of the Miller indices mapped in two dimensions. The surface Miller indices of four gaps, three gaps, and one gap, observed from <100>, <111>, and <110>, respectively, were plotted on the stereographic projection for each direction (Fig. S5). The stereographic projections at <100>, <111>, and <110> exhibit fourfold, threefold and twofold symmetry, respectively, which matches the 432 point group symmetry of the NP. The lack of threefold symmetry of <111> reflects some imperfection of a chiral gap located in the <111> direction. The chirality of a gap appeared in the Miller-index distribution of the regions of the stereographic projections (see, for example, the red region for S chirality in Fig. S6A) denoted $(\bar{l}\,h\,\bar{k})^S$ and $(h\,\bar{l}\,k)^S$ ($h>k>l>0$), which correspond to the upper and lower parts of the gap faces, respectively. The lack of $(\bar{l}\,h\,k)^R$ and $(h\,\bar{l}\,\bar{k})^R$ indicates that the mirror symmetry between the crystal facets is broken and that a high-Miller-index with S chirality is preferred for the measured nanocrystal. Strain near the surface and the internal strain distribution of the nanocrystal can provide a fingerprint of the interaction of the chiral molecules, in this case, the peptide, to induce a chiral concave morphology.

We analyzed the strain-field of 432 helicoid III with regard to the geometry and surface Miller indices. From the lattice displacement-field along the $\mathbf{Q}$ ($\mathbf{a_y}$) direction (Fig. S7), the strain-field was calculated using the $\mathbf{Q}$ component of the displacement gradient (Fig. 3A). Both tensile and compressive strain were observed on a scale of $10^{-3}$. In the surface strain distribution of a chiral gap facing $<10\bar{1}>$ perpendicular to $\mathbf{Q}$, the area indicated with a white dotted line in Fig. 3A, the compressive strain was observed at the concave edge, whereas tensile strain was observed at the side facets. For more detailed information, a contour map of the Miller-indices is shown. The Miller-index contour map of the gap (marked with a white dotted line) showed that the deepest edge consisted of $<00\bar{1}>$, <100>, and the high-Miller-indices connecting $<00\bar{1}>$ and <100>, whereas the side facets consisted of high-Miller-indices adjacent to {111}. This strain Miller-index relationship was confirmed by plotting the strain at each Miller-index on a $<10\bar{1}>$ stereographic projection (Fig. 3B).

We also discovered that the strain-field inside 432 helicoid III exhibited a symmetric pattern directly related to the concave gap structures and symmetry of the morphology.

The strain fields of the (100) and (10$\bar{1}$) planes with different depths of 432 helicoid III are shown in Fig. 3C. Compared with the cube (Fig. S8), which appeared to be almost strain-free, the strain of 432 helicoid III increased near the chiral gap. Interestingly, the strain-field showed twofold symmetry, that is, tensile strain distributed near the chiral gap at the top and bottom and compressive strain at the sides. This indicates that the strain-field developed along with the symmetry of 432 helicoid III, because twofold symmetry can only appear when one component of the strain-field is considered in fourfold symmetric (100) slices. If there was 432 symmetry in the strain-field, we would expect compressive strain in the $a_x$ and $a_z$ directions related to the tensile strain near the top deepest part of the chiral gap in the $a_y$ direction. This result again showed the correlation of the strain-field with the surface Miller-index, which also indicated a 432-symmetric distribution. Previous studies (*27*, *28*) have reported that adatoms on surface facets induce a strain-field in crystals. The building up of a strain-field inside 432 helicoid III can result from the steps and kinks as adatoms on the Au surface, which are presumably formed by the interactions of peptides.

The BCDI results can be directly used to better understand the chiral plasmonic response of a 432 helicoid III. Because the mechanical tensile and compressive strain can reconfigure the lattice constant of Au crystals, the corresponding dielectric constants of Au should be tuned accordingly (*29*). In particular, the intrinsic plasma frequency and core permittivity of Au were found to be adjusted by the atomic lattice constant (see the methods) (*30*, *31*). The spatially variant strains were substituted into the modified Drude-Sommerfeld model and the reconfigured dielectric constants according to the strains were calculated (Fig. S9). We then reconstructed 432 helicoid III with the spatially distributed dielectric constants (i.e., the refractive index, $n$), which was compared with a consistently mapped $n$ of 432 helicoid III (Fig. 4C). Given the range of the strain variations, the $n$ of 432 helicoid III can be spatially distributed with a scale of $10^{-3}$ (see Fig. S9 for more details).

This slight change in $n$ negligibly affects the far-field scattering of 432 helicoid III (Fig. S10). In particular, the extinction cross section could be changed by an order of magnitude of $10^{-16}$ owing to the mechanical strain. However, the resultant change in the near-field optical chirality of 432 helicoid III was found to be nontrivial. Over the last decade, the distinct concept of optical chirality (i.e., $C(\mathbf{r})= -\varepsilon_0\omega\text{Im}\{\mathbf{E}(\mathbf{r})\cdot\mathbf{H}^*(\mathbf{r})\}/2$) has been the gold standard for the nanophotonic-enabled ultrasensitive detection of molecular chirality (*13*, *14*, *32*). The locally induced plasmonic resonance can strongly enhance and squeeze the optical chirality into a deep-subwavelength-scale (referred to as a chiral hotspot). Consequently, this nanophotonic chiral hotspot in the near-field regime can address the dimensionally mismatched scale between the chiral molecules and the wavelength of interest and thus enormously improve the sensitivity of chiral molecular detection (*33*, *34*). In line with this, the deterministic prediction of the chiral hotspots at the surfaces of plasmonic NPs is crucial because the interactions between the chiral hotspots and molecules are prone to error from the photonic and molecular stochasticity.

At a wavelength of 630 nm, 432 helicoid III exhibited localized surface plasmon resonance (LSPR) (Fig. S11). In this regime, nontrivial corrections were made on the spatial distributions of the near-field enhancement of electric fields by the implementation of a strain-corrected $n$ of a 432 helicoid III NP (Fig. S11). As a result, the predictive distributions of optical chirality were corrected accordingly (Fig. 4B). Here, the cross-sectional distributions of optical chirality were taken at each position along the *x*-axis. Note that the spatial position and strength of the optical chirality (chiral hotspots) are heterogeneously distributed along the 3D surface of 432 helicoid III, because they depend on the LSPR mode. We then compared the accessible maximum distributions of optical chirality ($C_{max}(\pm)$) between a strain-induced $n$ and the constant $n$ of 432 helicoid III ($\Delta C_{max}(\pm)$). We found that $\Delta C_{max}(\pm)$ became nonnegligible (on the order of $10^{-1}$) at a specific position (at approximately +20 nm from the *x*-axis) (Fig. 4C), which could, in turn, leverage the accuracy of the predictive chiral hotspots and the resultant sensitivity of chiral molecular detection by nanophotonics.

Our results demonstrate that BCDI can be used to identify the 3D crystallographic structure of an NP with hidden concave surfaces. By utilizing the feature that BCDI is inherently related to the atomic structure, we successfully determined the surface-Miller-index and strain-field distributions of 432 helicoid III. An analysis of 432 helicoid III revealed that the chiral gaps consisted of high-Miller-index crystal-planes resulting from the enantioselective interactions of peptides and that a strain field correlated to the 432-symmetric morphology existed inside the NP. These results provide not only an understanding of the formation mechanisms but also tuning strategies for chiral morphologies and a blueprint for deep-subwavelength-scale chiral hotspots. Moreover, we anticipate that crystallographic identification with BCDI will serve as a generalized platform with in-situ/operando capability as well as an extended size regime and precise resolution. Real-time imaging facet identification during crystal growth (*35*) and catalytic reactions (*36–38*) can be applied in future research in catalytic, optical, and energy applications. The resolution limitation and the averaging effect for small facets are expected to be compensated for by the improvement in the diffraction-limited storage ring source (*39*).

**Acknowledgments:**
We acknowledge DESY (Hamburg, Germany), a member of the Helmholtz Association HGF, for the provision of experimental facilities. Parts of this research were carried out at P10, PETRA III. Beamtime was allocated for proposal I-20190530. Experiments at 9C, PLS-II were supported in part by MSIT and POSTECH. Use of the Advanced Photon Source (34ID-C) was supported by the Office of Basic Energy Science, under the Office of Science of the US Department of Energy (Contract No. DE-AC02-06CH11357).

**Funding:** This work was supported by the National Research Foundation of Korea grant NRF- 2021R1A3B1077076 (SC, SK, JK, HK), NRF-2017M3D1A1039377 (SWI, Y-CL, RMK, JHH, HK, KTN), NRF-2019R1A2C2004846 (J-HH, SL). This work was supported by the Technology Innovation Program funded By the Ministry of Trade, Industry & Energy (MOTIE, Korea) grant 20012390 (SWI, Y-CL, RMK, JHH, HK, KTN).

**Author contributions:** KTN and HK conceived and supervised this work. SC, SK, JK, MS, SYL, WC, RH, HK performed coherent X-ray diffraction imaging experiment. The imaging results and analysis were performed by SC. Samples were synthesized and characterized by SWI, Y-CL, RMK, JHH under the supervision of NKT. Simulation analysis of strain and chiroptical characteristics were carried out by J-HH and SL. SC, SWI, SL, KTN, HK wrote the manuscript. All authors discussed the results and commented on the manuscript.

   Conceptualization: SC, SWI, KTN, HK

   Investigation: SC, SWI, J-HH, SK, JK, MS, SYL, WC, RH

   Visualization: SC, SWI

   Formal analysis: SC, J-HH, Y-CL, RMK, JHH

   Resources: MS, SYL, WC, RH

   Supervision: KTN, HK

Writing – original draft: SC, SWI, SL, KTN, HK

Writing – review & editing: SC, SWI, SL, KTN, HK

**Competing interests:** Authors declare that they have no competing interests.

**Data and materials availability:** All data are available in the main text or the supplementary materials.

## Supplementary Materials

Materials and Methods

Supplementary Text

Figs. S1 to S13

Tables S1

References (*40–46*)

Movies S1

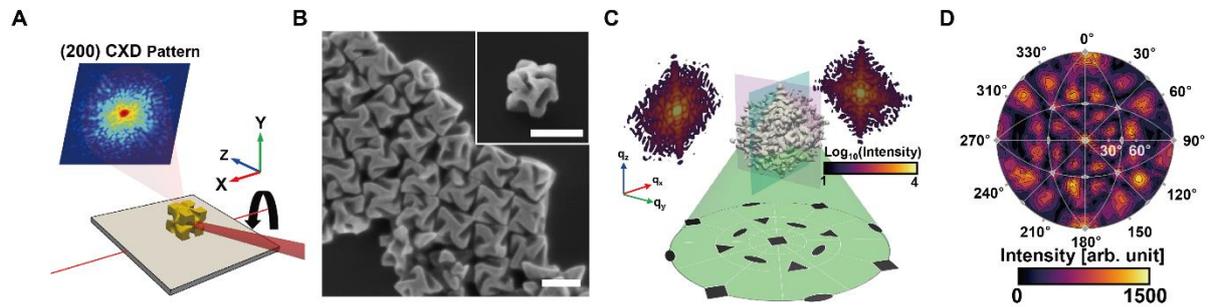

**Fig. 1. Bragg coherent X-ray diffraction imaging and CXD pattern-based analysis of chiral nanoparticles.** (**A**) Schematic illustration of the BCDI of 432 helicoid III. A focused coherent X-ray was illuminated on 432 helicoid III. The CXD patterns were measured at the Au (200) Bragg reflection with sample rotation along the rocking curve. (**B**) SEM image of 432 helicoid III NPs showing the cubic outline and chiral gap structures. (**C**) Measured 3D CXD patterns of 432 helicoid III. (**D**) Calculated stereographic pole figure projection of the CXD pattern. The {100}, {111}, and {110} points indicated fourfold, threefold, and twofold symmetry, respectively.

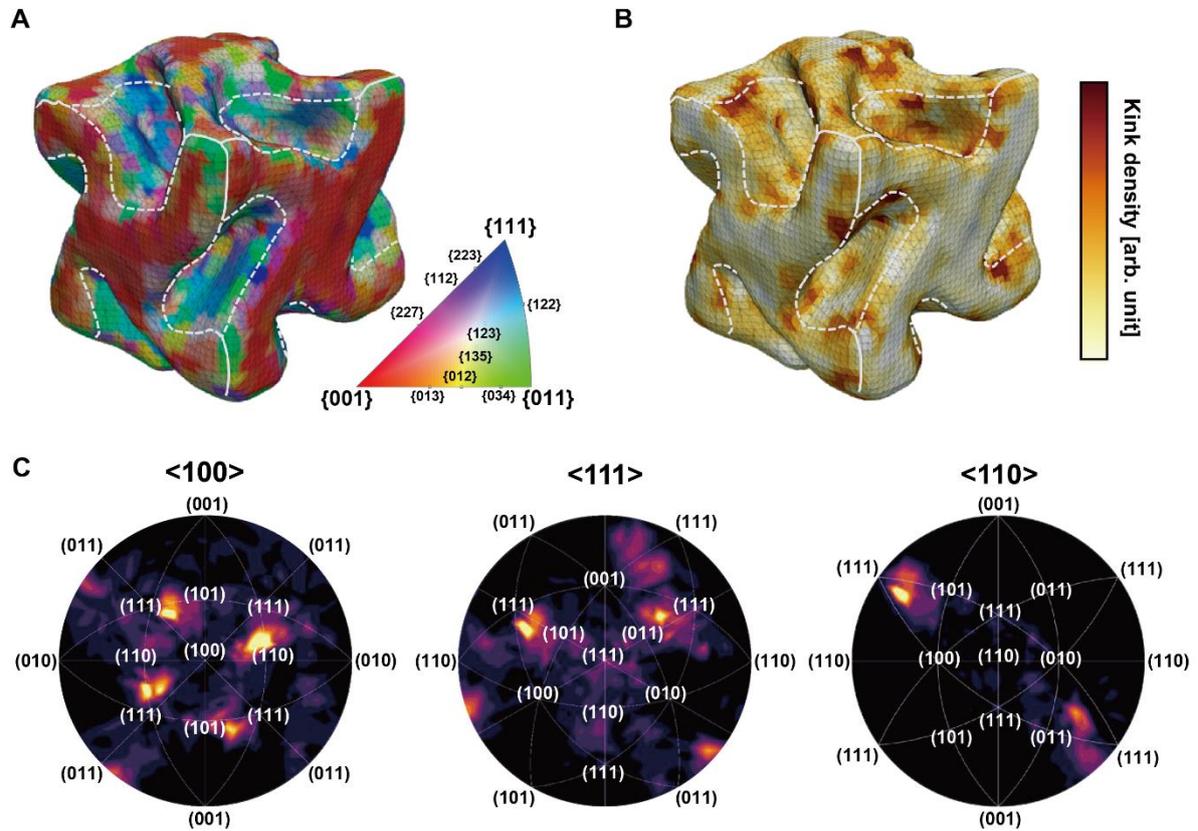

**Fig. 2. Surface Miller-index analysis of 432 helicoid III.** (**A**) Surface Miller-indices and (**B**) kink density of 432 helicoid III. The white dashed lines indicate the boundaries between the cubic surface (100) and gap regions. The (100), (110), and (111) planes are represented by red, green, and blue, respectively, with the intermediate Miller indices expressed as a color code. The kink density was calculated from the extracted surface Miller indices. (**C**) Distributions of the surface Miller-index on the crystallographic stereographic projection along the <100>, <111>, and <110> directions. The surface area of each facet of four gaps, three gaps, and one gap is expressed as the brightness on the respective stereographic projections of the Miller indices. Each projection showed approximately fourfold, threefold, and twofold symmetry, matching the point group symmetry of 432 helicoid III.

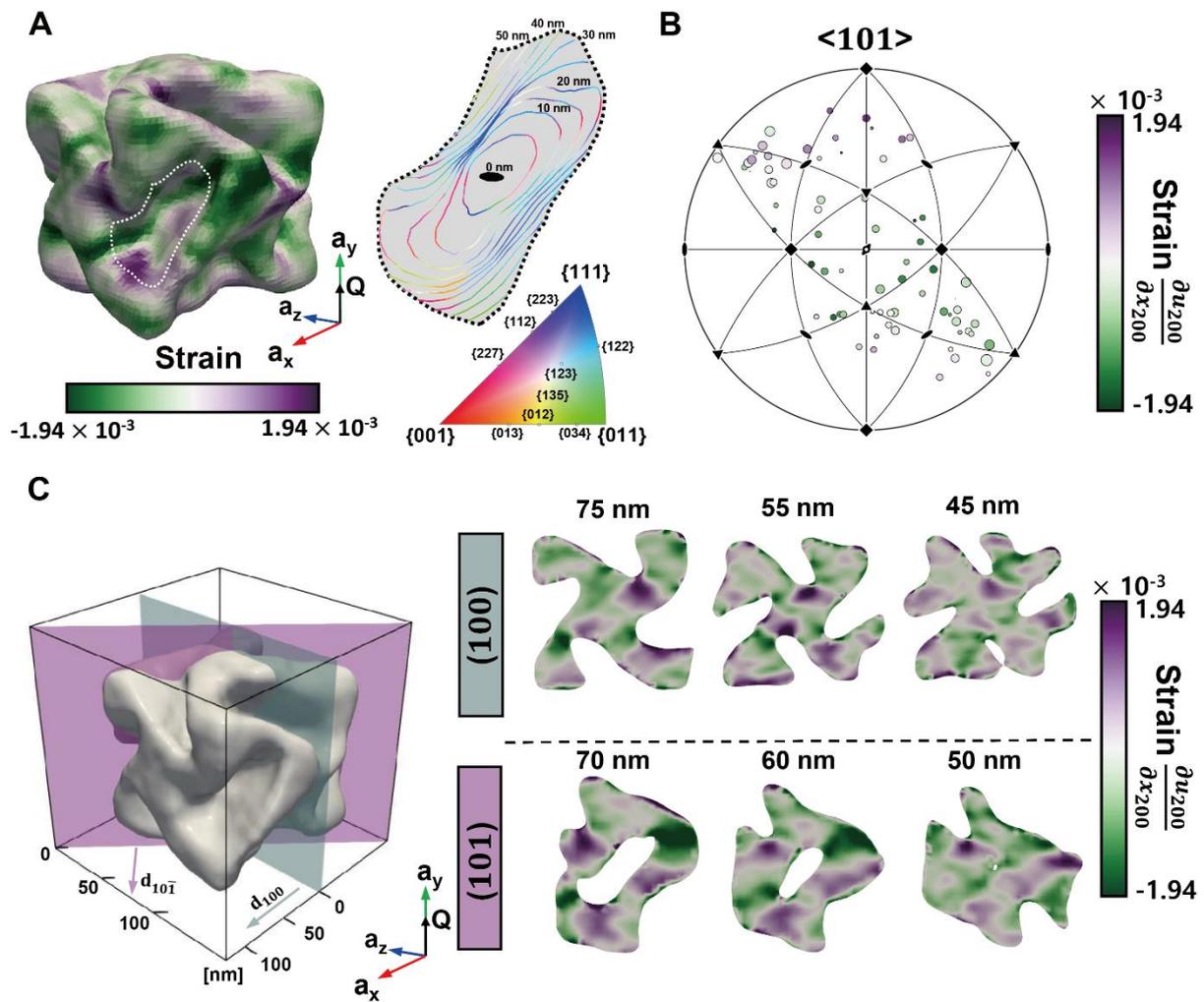

**Fig. 3. External and internal strain-field distributions induced by geometric symmetry.** **(A)** Surface strain of 432 helicoid III and detailed contour map of the Miller indices of the chiral gap facing [10$\bar{1}$] (indicated with a white dotted line). **(B)** Stereographic projection of the surface strain distribution for the [10$\bar{1}$] chiral gap in (A). The size of each circle indicates the area of the corresponding Miller-index. **(C)** Internal strain-field distribution of 432 helicoid III. The (100) and (10$\bar{1}$) slices of the strain-field at different distances from the center are shown. The strain was concentrated in the deep parts of the chiral gap, and its distribution showed symmetry that indicates 432 symmetry in the strain-field.

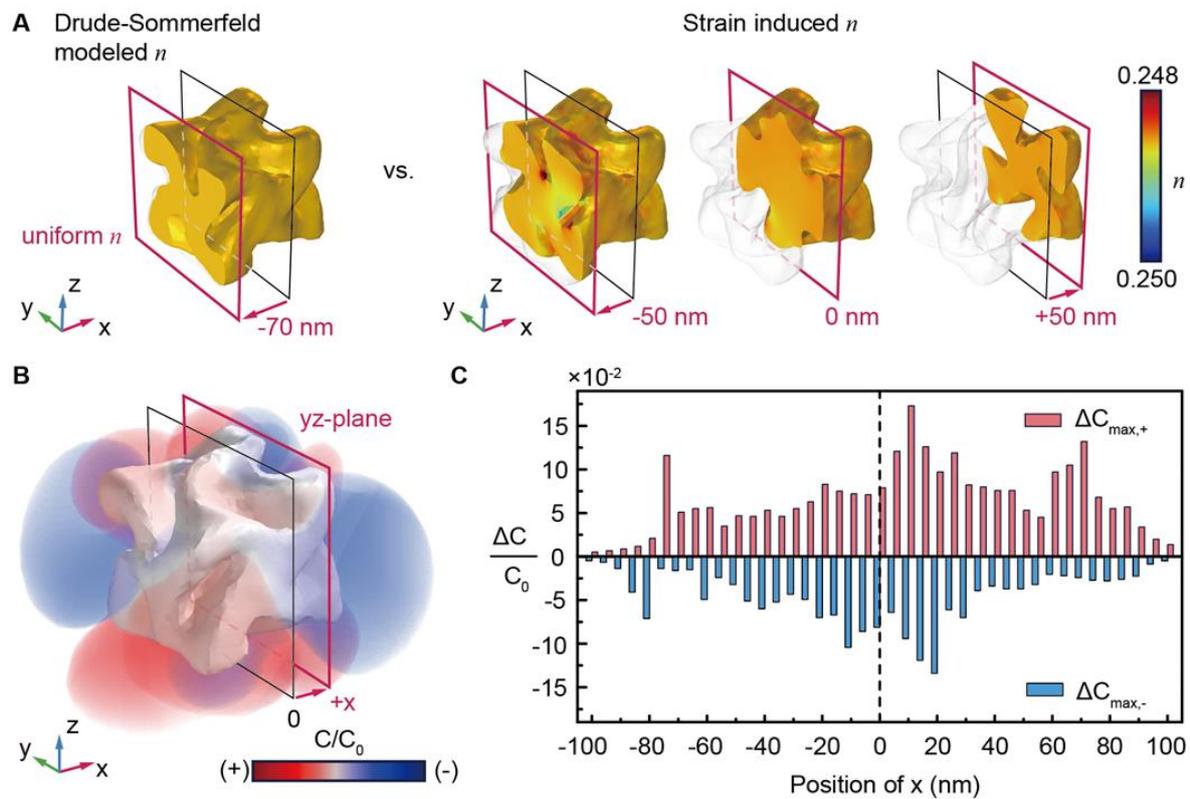

**Fig. 4. Simulation analysis of the strain and chiroptical characteristics of 432 helicoid III.** (**A**) Three-dimensional spatial color maps of the calculated refractive index based on the Drude-Sommerfeld and strain-induced model at 630 nm. (**B**) Three-dimensional optical chirality enhancement distributions. (**C**) Optical chirality difference between the Drude-Sommerfeld and strain-induced models.

# Supplementary Materials for

## Strain and Crystallographic Identification of the Helically Concaved Surfaces of Nanoparticles

Sungwook Choi[1]†, Sang Won Im[2]†, Ji-Hyeok Huh[3], Sungwon Kim[1], Jaeseung Kim[1], Yae-Chan Lim[2], Ryeong Myeong Kim[2], Jeong Hyun Han[2], Hyeohn Kim[2], Michael Sprung[4], Su Yong Lee[5], Wonsuk Cha[6], Ross Harder[6], Seungwoo Lee[3,7], Ki Tae Nam[2]*, Hyunjung Kim[1]*

[1]Department of Physics, Sogang University; Seoul 04107, Korea.
[2]Department of Materials Science and Engineering, Seoul National University; Seoul 08826, Korea
[3]KU-KIST Graduate School of Converging Science & Technology, Korea University, Seoul 02481, Korea
[4]Deutsches Elektronen-Synchrotron (DESY), Hamburg 22607, Germany
[5]Pohang accelerator laboratory, Pohang 37673, Korea
[6]Advanced Photon Source, Argonne National Laboratory, Argonne, Illinois 60439, USA
[7]Department of Integrative Energy Engineering and KU Photonics Center, Korea University, Seoul 02481, Korea
†These authors contributed equally to this work
*Corresponding author. Email: nkitae@snu.ac.kr, hkim@sogang.ac.kr

**This PDF file includes:**

    Materials and Methods
    Supplementary Text
    Figs. S1 to S13
    Tables S1
    Captions for Movies S1
    References and Notes

**Other Supplementary Materials for this manuscript include the following:**

    Movies S1

**Materials and Methods**

Chemicals

Hexadecyltrimethylammonium bromide (CTAB, 99%), hexadecyltrimethylammonium chloride (CTAC, 98%), L-ascorbic acid (99%), tetrachloroauric(III) trihydrate ($HAuCl_4 \cdot 3H_2O$, 99.9%), L-glutathione (98%), sodium tetrahydridoborate ($NaBH_4$, 99%), and potassium iodide (KI, 99.5%) were purchased from Sigma-Aldrich. High-purity deionized water (18.2 $M\Omega\ cm^{-1}$) was used to prepare aqueous solution.

Synthesis of 432 helicoid III

432 helicoid III was synthesized as reported previously (*1*). Spherical seed (~2.5 nm) was prepared by mixing 250 μl of 10 mM $HAuCl_4$ and 10 ml of 100 mM CTAC, followed by injecting 250 μl of ice-cold $NaBH_4$. The solution was kept at 30°C for 2 h. The growth solution of octahedral seed was prepared by mixing 250 μl of 10 mM $HAuCl_4$, 9.5 ml of 100 mM CTAC, 5 μl of 10 mM KI, and 220 μl of 40 mM L-ascorbic acid. 55 μl of diluted spherical seed solution was injected and kept at 30°C for 15 min. The solution was centrifuged (6708*g*, 150 s) twice and redispersed in 1 mM CTAB solution. The growth solution of 432 helicoid III was prepared by mixing 3.95 ml of deionized water, 800 μl of 100 mM CTAB, 100 μl of 10 mM $HAuCl_4$, 475 μl of 100 mM L-ascorbic acid, and 5 μl of 5 mM L-glutathione. 50 μl of diluted octahedral seed solution was injected and kept at 30°C for 2 h. The solution was centrifuged (1677*g*, 60 s) twice and redispersed in 1 mM CTAB solution. Cube nanoparticle was prepared by the growth of octahedral seed in growth solution without L-glutathione.

Preparation of nanoparticle-loaded substrate

432 helicoid III and cube nanoparticle solutions were concentrated and redispersed in 10 μM CTAB solution. The solutions were spin-coated on a square silicon substrate with a side length of 1.5 cm. The morphology of nanoparticles was measured by scanning electron microscopy (SEM). 100 nm of $SiO_2$ protection layer was coated on the substrate by E-beam evaporation.

BCDI experimental details

BCDI experiment for 432 helicoid III nanoparticles was conducted at P10 beamline, PETRA III (DESY, Germany). Using focused coherent X-ray (1×1.21 μm$^2$) by the compound refractive lens with 8.84keV photon energy, we measured CXD patterns in the vicinity of Au (200) Bragg peak. Each CXD pattern was collected by a 2D detector (Eiger 4M with 75μm pixel size) with rotation of nanoparticles on the goniometer along the rocking curve. , The sample to detector distance was 1.84 m. For the purpose of facilitating high spatial resolution, a wide range (-1.2° ~ 1.2°) of rocking scans with 0.01° step size was measured.  BCDI for cube gold nanoparticles was measured at 9C beamline, PLS-II, Korea. The measurement was done with a focused X-ray by KB-mirror (6.8× 13.1 μm$^2$) with 8 keV photon energy. CXD patterns were also measured in Au (200) geometry by 2D detector (Timepix, 55μm pixel size). The sample to detector distance was 0.73 m. The range of the rocking scan was 1.2° with a 0.01° step size.

In 34-ID-C beamline, APS, USA, a preliminary experiment was conducted. The measurements were done with focused X-ray by KB-mirror (0.45×1.3 μm$^2$) with 10 keV photon energy. CXD patterns were also measured in both Au (200) and (111) geometry by a 2D detector (Timepix, 55μm pixel size). For establishing feasible sample preparation conditions, various sample to detector distances and scan ranges were examined.

Phase retrieval process

Total 13 successive scans for the same 432 helicoid III nanoparticle were accumulated to improve signal-to-noise ratio. After summing up scans, the diffraction pattern was binned typically (3 by 3) due to the larger than necessary oversampling of the reciprocal space coherent interference patterns caused by the long sample to detector distance. Phase retrieval process (*25*) was conducted with combinations of ER and HIO algorithms (20:180) and the total number of iterations was 620. Initial support was square with a random phase. All trials of reconstructions yield similar results. The phase retrieval twin solution of 432 helicoid III were determined by the handedness of chiral morphology. The calculation was accelerated by GPU-based computing (NVIDIA TESLA K20). Phase retrieval for cube nanoparticle was conducted by the same procedure without accumulation in the measurements and binning of the measured data.

**Supplementary Text**

Stereographic projection and inverse pole figure plot of CXD pattern

To generate stereographic projection of CXD patterns (Fig. 1G and 1H) (*24*), the measured pattern was transformed into reciprocal space that conjugated with the basis of sample space. During transformation, the reconstructed phase was treated as zero for the purpose of excluding pattern distortion that originated from lattice strain. For this reason, the transformed CXD pattern became ideally symmetric, and we only consider the upper hemisphere of the pattern. Because there is a trade-off relation between the solid angle of fringe and intensity, radius (r) and width of hemisphere ($\Delta r$) was selected at near $3^{rd} \sim 4^{th}$ order fringe to achieve both precise angular information and enough signal-to-noise ratio. For pattern of 432 helicoid III, r = 0.0137 Å$^{-1}$ and $\Delta r$ = 0.0016 Å$^{-1}$ was selected. For cube nanoparticle, r = 0.0203 Å$^{-1}$ and $\Delta r$ = 0.0016 Å$^{-1}$ was selected. Both stereographic projections were plotted in (100) direction. To present primary low-Miller-index, the symmetry symbol was overlaid. Indexing of stereographic projection was performed by MTEX toolbox (*40*).

Detailed calculation procedure of coordinate transformation

The transformation of the 3D image is based on the basis transformation of space coordinates (*21*). The space coordinate of the 3D image was transformed for a matching basis with direct lattice vectors. The process of the coordinate transformation is shown in Fig. S3A which consisted of two Rodrigues rotation matrices ($R_1$ and $R_2$). The first step was the calculation of $R_1$, which transforms the basis of lab space basis (I) into Q-corrected space basis (II). Rotation matrix $R_1$ that is represented in the figure between (I) and (II) makes wavevector transfer (**Q**) becomes a new +**y**′ basis vector of (II). In contrast to the first transformation $R_1$, the calculation of $R_2$ was not based on the actual posture of a nanoparticle. To align the surface normal vector (**N**) to one of the direct lattice vectors ($\mathbf{a_z}$) by rotating the 3D image around +**y**′, we introduced terrace area ($S_T$), where the total terrace area of crystal surface in the Terrace-Step-Kink (TSK) model (*16*), as a fitting parameter. In the TSK model, the area proportion of each facet with Miller-index (*hkl*) is decomposed into micro-facets as below.

$$(h\ k\ l) = u \times (100) + v \times (110) + w \times (111)$$

where Miller-index (*hkl*) is sorted from largest to smallest ($h \geq k \geq l \geq 0$). After decomposition of the surface, terrace area $S_T$ was calculated for each rotation angle along with **Q** that is parallel to the **+y′**. $S_T$ is defined as

$$S_T = \Sigma A_{MF}^{(uvw)} \times \frac{(u\ v\ w)}{|(u\ v\ w)|} \cdot (1\ 0\ 0)$$

where $A_{MF}^{(uvw)}$ is an area of decomposed facet. Figure S3B shows the fitting results of the measured $S_T$. To consider only flat outer facet, fitting for 432 helicoid III considered only near the surface segment of the nanoparticle that was presumably considered as (100) surface. The terrace area was calculated with a range of -30°~30° and 2° step size starting from the arbitral **Q**-normal vector in **Q**-corrected space. The fitting formula was $S_T = a_1 \sin(\omega\theta) + b_1 \cos(\omega\theta) + c$ to fit the area that was changed in the area of two major surfaces perpendicular to each other. $R_2$ was selected at the maximum point of the $S_T$ that maximizes the area of the "red" facet, i.e. {100} facet. To visualize the process of determination of rotation angle, movie. S1 was presented in a supplementary video1. Final transformed 3D images are shown in (III) that all direct lattice vectors are parallel to the basis of (III). For this reason, the calculated normal vector of the surface in the sample space is equivalent to the normalized Miller-index of the surface. Rotation angle along **Q** was determined by fitting of terrace area ($S_T$) where total terrace area of crystal surface in the TSK model (*26*).

Iterative surface segmentation and fitting

Figure S4 shows the schematic flow chart and segmentation results during the early stage of iteration. For every generation, all point clouds were divided into two segments by the K-means segmentation method that minimizes the sum of squared distance within all segments (*40*). As shown in Fig. S4A, termination of segmentation was determined by the root-mean-square (RMS) roughness of the segmented surface. Because a surface whose RMS roughness is less than pixel resolution can be considered a perfectly flat surface for BCDI spatial resolution, segmentation was terminated if RMS roughness became less than 1 pixel resolution. Fig. S4 (B, C, and D) shows early states of segmentation. Each color represents different segments. After iterative segmentation, total 1587 flat surfaces were resolved. Each point cloud in surfaces was fitted by the singular value decomposition (SVD) method (*42*). Using the SVD method, each point cloud was plane fitted and Miller-index was extracted from the normal vector of the fitted surface. Color code for representation of crystallographic orientation use TSL color code in MTEX toolbox that converts orientation into color in HSV color-space. These colors were mapped on the iso-surface of imaged nanoparticles (Fig. 2A and Fig. S12A). All series of image processing were conducted by in-house MATLAB code.

Calculation of kink density

The kink density of high-Miller-index surface of face-centered-cubic (FCC) gold crystal was evaluated based on TSK model (*26*). The atomic surface structure of an ideal crystal facet with Miller-index (*hkl*) can be described by the composition of low-Miller-index micro facets of (111), (100), and (110). Assuming that Miller-indices are scaled to the smallest possible integers and $h \geq k \geq l \geq 0$, Miller-index can be decomposed to

$$(hkl) = l(111) + (k - l)(110) + (h - k)(100)$$

and for the FCC crystal lattice, the number of unit cells is derived as

$$n_{hkl} : n_{111} : n_{110} : n_{100} = \text{p}: 4l : 2(k - l) : 2(h - k)$$

where $p$ is 2 when $h, k, l$ are not all odd, and 4 when $h, k, l$ are all odd. The unit cell area for the Miller-index ($hkl$) is

$$A_{hkl} = \frac{a^2}{2}\sqrt{h^2 + k^2 + l^2} \quad \text{for h, k, l not all odd,}$$
$$A_{hkl} = \frac{a^2}{4}\sqrt{h^2 + k^2 + l^2} \quad \text{for h, k, l all odd,}$$

where $a$ is the number of kinks is equal to the number of least microfacet. Therefore, the number of kinks for the area can be derived as

$$d_{hkl}^{kink} = \frac{\min\{4l, 2(k-l), 2(h-k)\}}{pA_{hkl}} = \frac{\min\{4l, 2(k-l), 2(h-k)\}}{a^2\sqrt{h^2 + k^2 + l^2}}$$

The calculated R/S kink density and kink density on the stereographic triangle is illustrated in Fig. S6.

Consistency of the strain-field distribution

We verified that another 432 helicoid III NP showed a symmetric strain-field distribution consistent with the main results (Fig. 3C and Fig. S13). There are tensile strain distributions near the chiral gap at the top and bottom and compressive strain at the sides. Another NP was measured under the same conditions, with the exception of the rocking range (-0.6° ~ 0.6°).

The effect of strain on the dielectric function

We modified Drude-Sommerfeld dielectric function to obtain the dielectric function of the Au crystals, reconfigured by a mechanical strain (*29*). Since the lattice of Au atoms can be tuned due to the strain, the dielectric function, which represents the bulk properties of light-matter interaction, should be accordingly changed. This mechanical modification mainly alters the plasma frequency ($\omega_p$) and $\varepsilon_{core}$, which are mainly determined by the positive ion cores of the atom. To do this, we began with the modified bulk Drude-Sommerfeld dielectric function that can reflect the effect of nanosized materials (*43*). The corresponding Drude-Sommerfeld dielectric function is as follows:

$$\varepsilon_{bulk}(\omega) = \varepsilon_{core}(\omega) - \frac{\omega_p^2}{\omega^2 + i\omega\left(\gamma_0 + K\frac{v_F}{l_{eff}}\right)}$$

where $\varepsilon_{core}$ corresponds to electric permittivity representing the bound electrons; $\omega_p$ is plasma frequency; $\gamma_0$ indicates the damping constant related to the collision of conduction electrons within the bulk materials. $\omega_p$ is defined by $\sqrt{ne^2/\varepsilon_0 m_{eff}}$, where $n$, $m_{eff}$, $e$, $\varepsilon_0$ are the electron density, effective mass, elementary charge, and the permittivity of vacuum, respectively.

The main difference between typical and modified Drude-Sommerfeld is the additional term $Kv_F/l_{eff}$ in the bracket. This term rationalizes the possibly enhanced collision rate, particularly for the increased ratio of the surface area to volume (e.g., Au nanoparticle (NP)). Herein, $v_F$ and $l_{eff}$ are the Fermi velocity and the reduced effective mean free path, respectively: $l_{eff}$ is defined by $4V/S$, where $V$ is the volume and $S$ is the surface area of the NP. Lastly, $K$ is a dimensionless parameter.

The parameters in Table S1 are applied to the modified bulk dielectric function to get the bulk dielectric constants of the nanosized Au crystal. (*44, 45*)

Lastly, we modified $\omega_p$ to account for the electron density modulation, which is driven by the applied mechanical strain. The modified plasma frequency, $\omega_p^{strain}$, can be written as (*30*)

$$\omega_p^{strain} = \sqrt{\frac{4e^2}{\varepsilon_0 m_{eff} a_{strain}^3}}$$

where $a_{strain}$ is the deformed lattice constant. In this work, we reconfigured $a_{strain}$ according to the numerically calculated curvature-induced strain tensor, shown in Fig. 4B. Furthermore, the plugged mechanical strain affects the dielectric function for the bound electrons $\varepsilon_{core}$. Thus, strain-induced $\varepsilon_{core}$ can be rewritten as $\varepsilon_{core}^{strain}$ and this relation is as follows: (*31*)

$$\varepsilon_{core}^{strain}(\omega) = \frac{\varepsilon_{core} + 2 + 2\nu(\varepsilon_{core} - 1)}{\varepsilon_{core} + 2 - \nu(\varepsilon_{core} - 1)}$$

where $\nu$ is the strain-tuned lattice constant ($\nu = (a_0/a_{strain})^3$, where $a_0$ is the lattice constant). Note that we neglected the quantum effects, which can be justified by the relatively large size of the Au NP (*46*). According to the following results, the strain-induced dielectric function can be written as

$$\varepsilon_{bulk}^{strain}(\omega) = \varepsilon_{core}^{strain}(\omega) - \frac{\omega_p^{strain\,2}}{\omega^2 + i\omega\left(\gamma_0 + K\frac{v_F}{l_{eff}}\right)}$$

As shown in Fig. S11, not only the strain-induced dielectric function (relative permittivity) (Fig. S11A), but also the strain-induced complex refractive index was calculated based on the curvature-induced strain through the strain-induced dielectric function (Fig. S11B).

Numerical simulation for the strain-induced optical properties

To investigate the effect of strain on both the far-field and near-field optical properties of the Au 432 helicoid III, we carried out the numerical calculation by using a commercial finite-element solver (COMSOL Multiphysics). The simulation domain was encapsulated by the perfectly matched layers (PML). We imported the reconstructed NP model obtained from CDI, and linearly polarized light was illuminated. The dielectric function of Au was embedded by adapting the formula derived from the previous section (i.e., Drude-Sommerfeld and strain-induced model), and the water was used as the host medium with a refractive index of 1.33. The far-field characteristics of 432 helicoid III were estimated through the extinction cross-section (ECS), which is the summation of scattering cross-section (SCS) and absorption cross-section (ACS). Herein, the SCS was derived by the surface integration of scattered Poynting vector passing the envelope, while ACS was obtained from the integration of the electromagnetic power loss density inside the nanoparticle (Fig. S10B). The difference in ECS between the Drude-Sommerfeld model and the strain-induced model was also calculated in Fig. S10C.

In addition, as shown in Fig. 4 and Fig. S11, we analyze the near-field optical properties. The optical chirality $C$ is calculated using the following equation.

$$C(\mathbf{r}') = \frac{\varepsilon_0 \omega}{2} Im\{\mathbf{E}_{sca}(\mathbf{r}') \cdot \mathbf{H}_{sca}(\mathbf{r}')^*\}$$

where $C(\mathbf{r}')$ of the scattered field at the position $\mathbf{r} = \mathbf{r}'$. $\varepsilon_0$ and $\omega$ correspond to the permittivity at the vacuum and angular frequency of incidence light, respectively. $\mathbf{E}_{sca}$, and $\mathbf{H}_{sca}$ indicate scattered electric and magnetic fields, respectively. Both three-dimensional $C$ and $\mathbf{E}_{sca}$ distributions are displayed through the multiple slice plots calculated from the numerical simulation data (Fig. 4D and Fig. S8A).

**Fig. S1.**

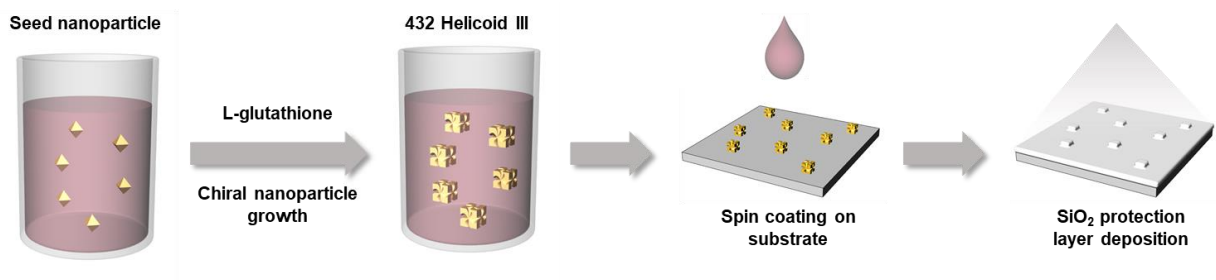

**Fig. S1. Preparation of 432 helicoid III nanoparticles.** Schematic description of 432 helicoid III nanoparticle sample preparation.

**Fig. S2**

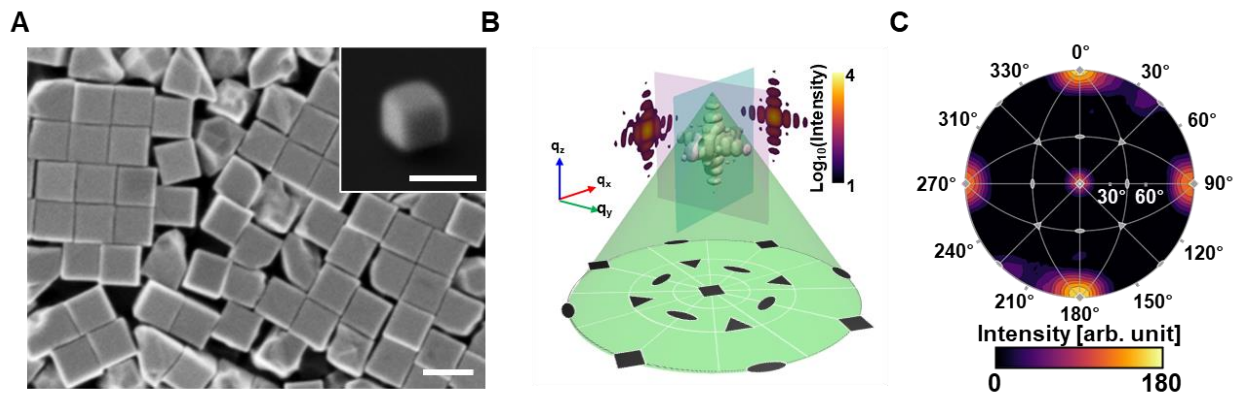

**Fig. S2. Scanning electron microscopy (SEM) image and stereographic pole figure projection analysis for cube NP.** **(A)** Scanning electron microscopy (SEM) image of cube nanoparticles. **(B)** Measured three-dimensional CXD patterns cube. The stereographic pole figure projection (green) is calculated for the upper slice of the hemisphere. **(C)** Stereographic pole figure projection of the CXD pattern of the cube. The {100}, {111}, and {110} points are indicated with fourfold, threefold, and twofold symmetry, respectively.

**Fig. S3.**

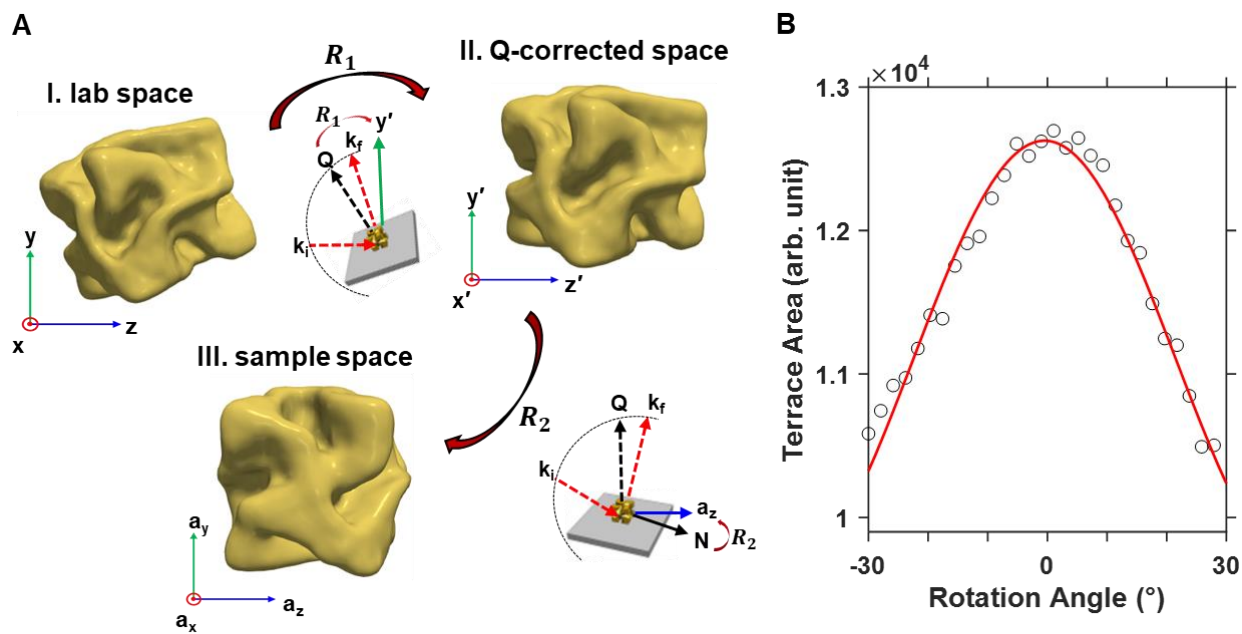

**Fig. S3 Coordinate transformation process of 432 helicoid III.** (**A**) The coordinate transformation process from the lab space to the sample space for 432 helicoid III. 3D images in lab space (I) are transformed into images in sample space (III) using two Rodrigues' rotation matrix $R_1$ (I to II) and $R_2$ (II to III). The figures between (I-III) show the relation between basis and other geometrical indicators. (**B**) Fitting of the terrace area $S_T$ for calculation of transformation matrix $\mathbf{R_2}$ for 432 helicoid III.

**Fig. S4.**

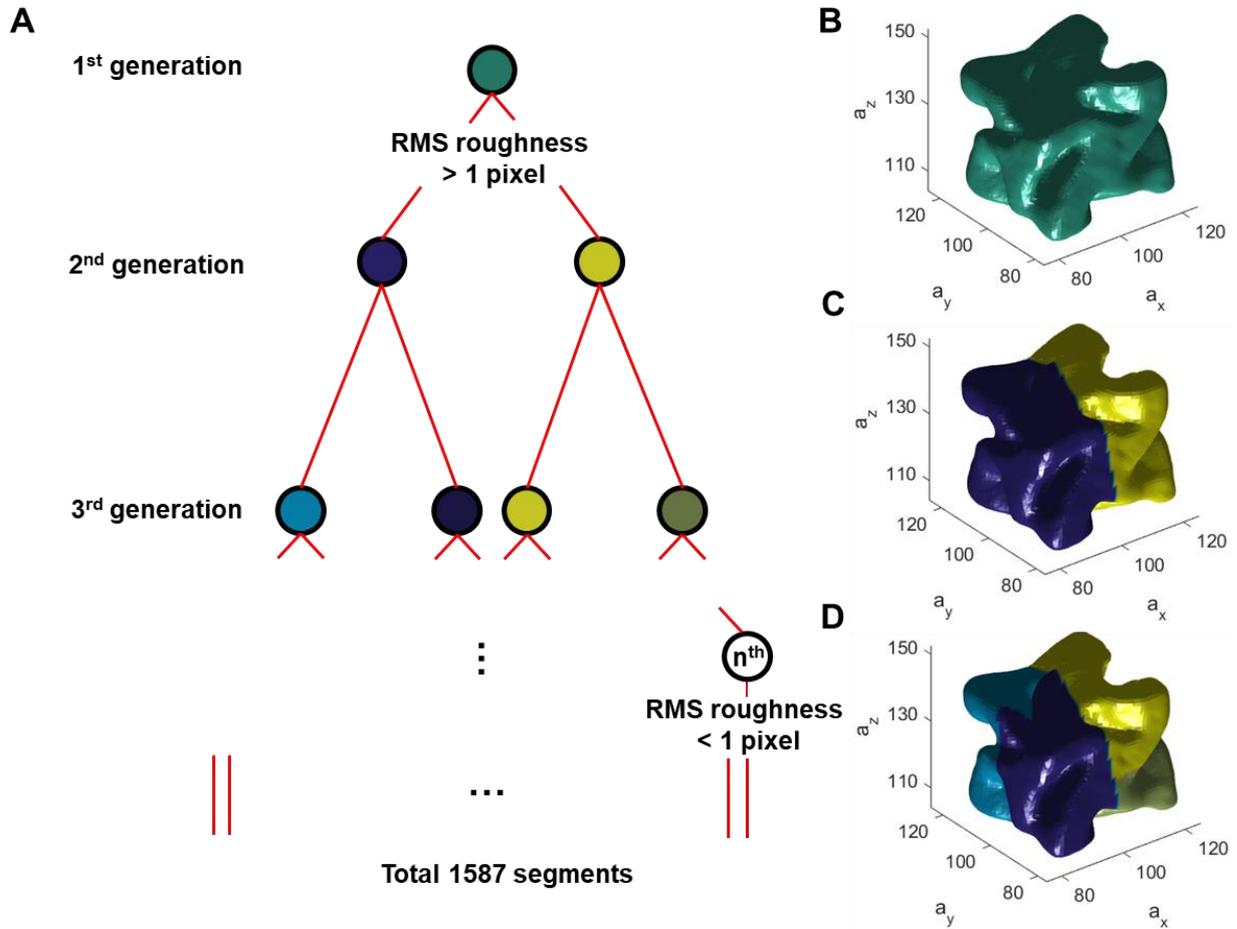

**Fig. S4. Iterative K-means surface segmentation of 432 Helicoid III nanoparticle. (A)** Schematic flow chart of the iterative segmentation algorithm. Each segment is divided into two sub-segments by the K-means clustering method unless root-mean-square roughness become less than 1 pixel (4.175 nm). **(B-D)** First three generations of the iterative surface segmentation algorithm. Each subsegment is represented with a different color.

**Fig. S5.**

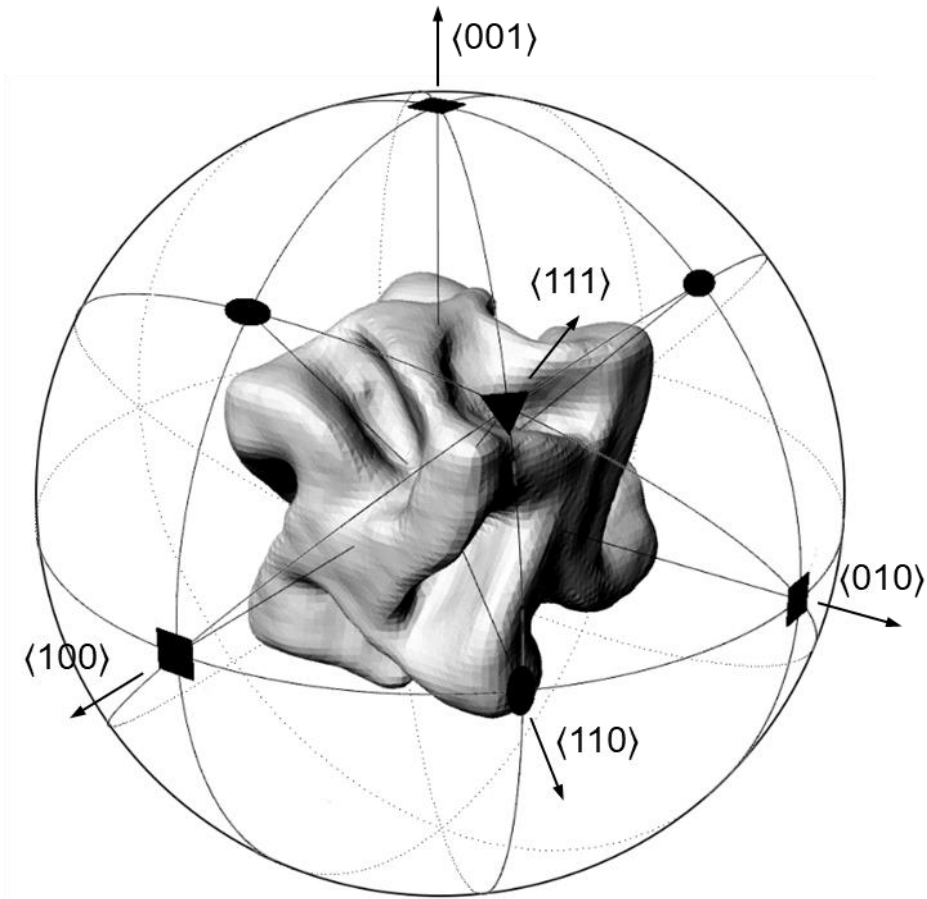

**Fig. S5 Reconstructed image of 432 helicoid III.** Reconstructed 3D geometry of 432 helicoid III nanoparticles. Symmetry symbols on the sphere indicate fourfold, threefold, and twofold symmetry axes of 432 point group symmetry.

**Fig. S6**

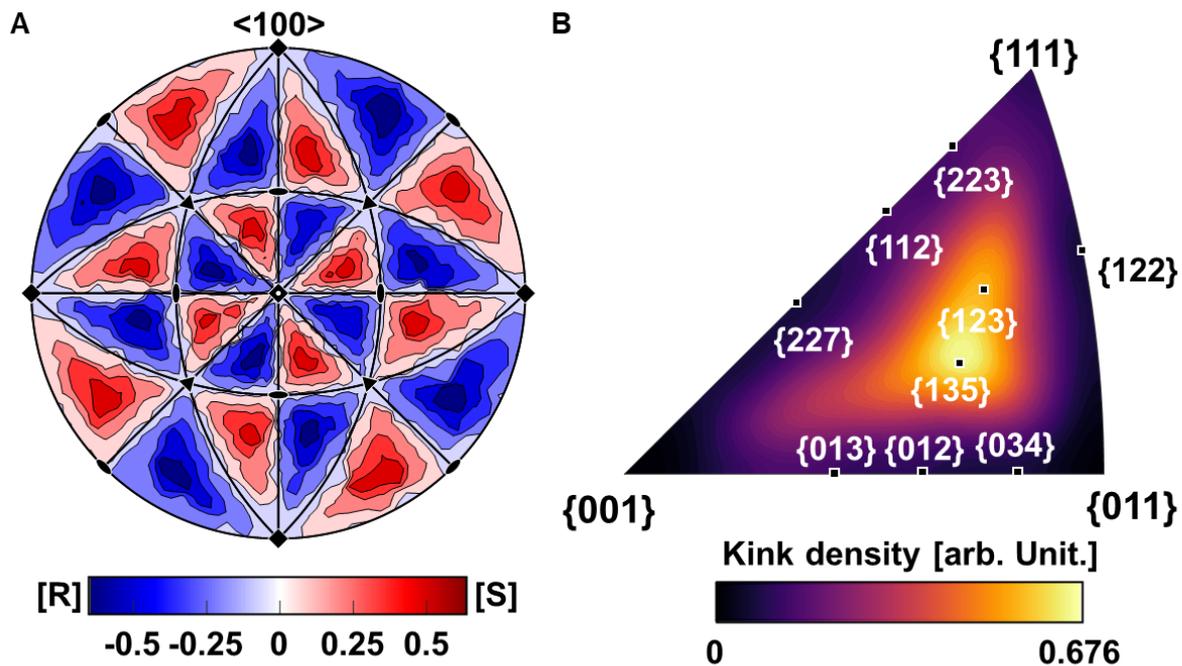

**Fig. S6. Stereographic description of R/S chirality and kink density.** (**A**) Stereographic projection of R/S kink density with symmetry symbols. (**B**) Kink density on the stereographic triangle.

**Fig. S7.**

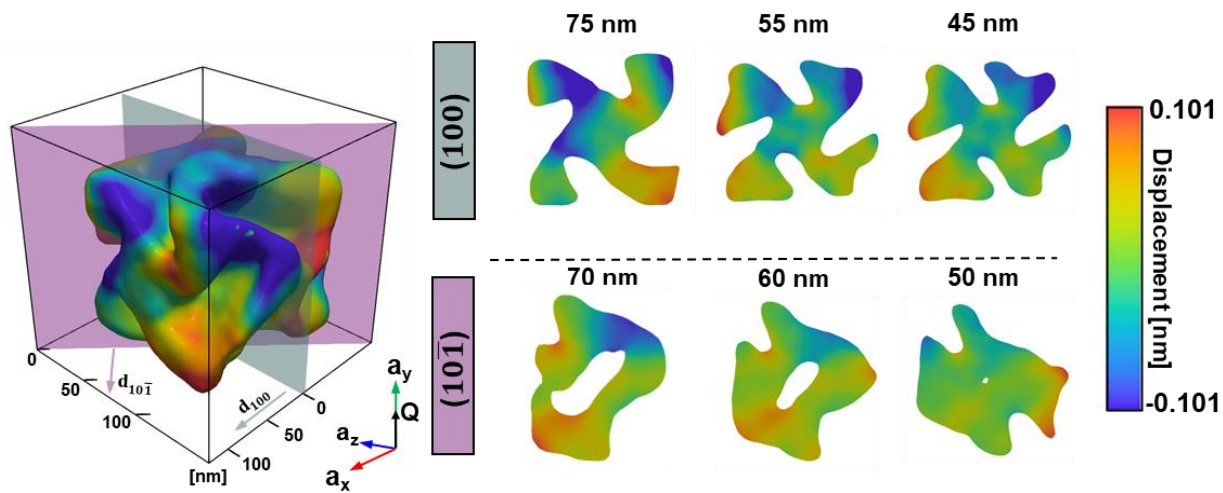

**Fig. S7. Displacement field of 432 helicoid III.** Lattice displacement-field slices of the 432 helicoid III for (100) and (10$\bar{1}$) planes. Each distance from the center is indicated at the outline of the nanoparticle and slices. (Same slices in Fig. 4A).

**Fig. S8.**

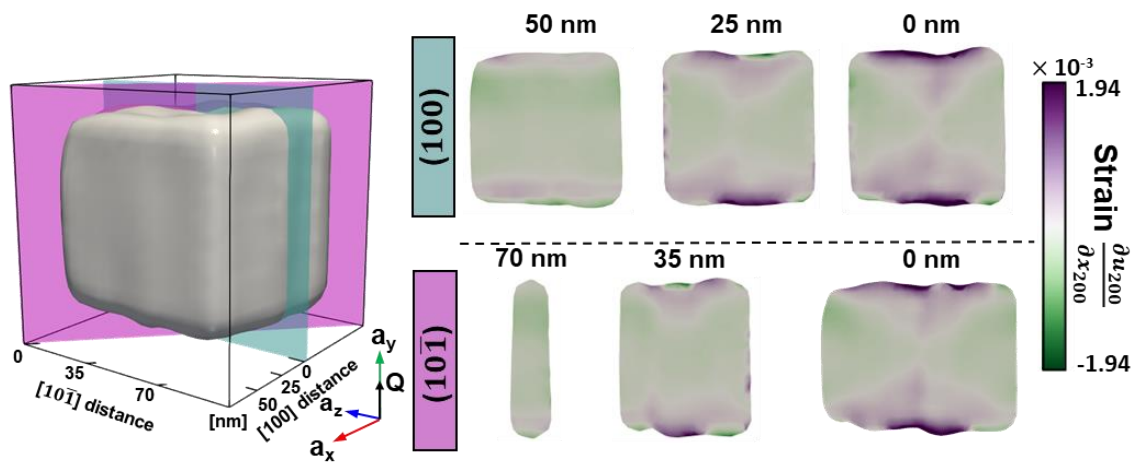

**Fig. S8. Strain slices of cube NP.** Strain slices of the cube NP from the center for (100) and (10$\bar{1}$) planes. Each distance is indicated at the outline of the nanoparticle and slices.

**Fig. S9.**

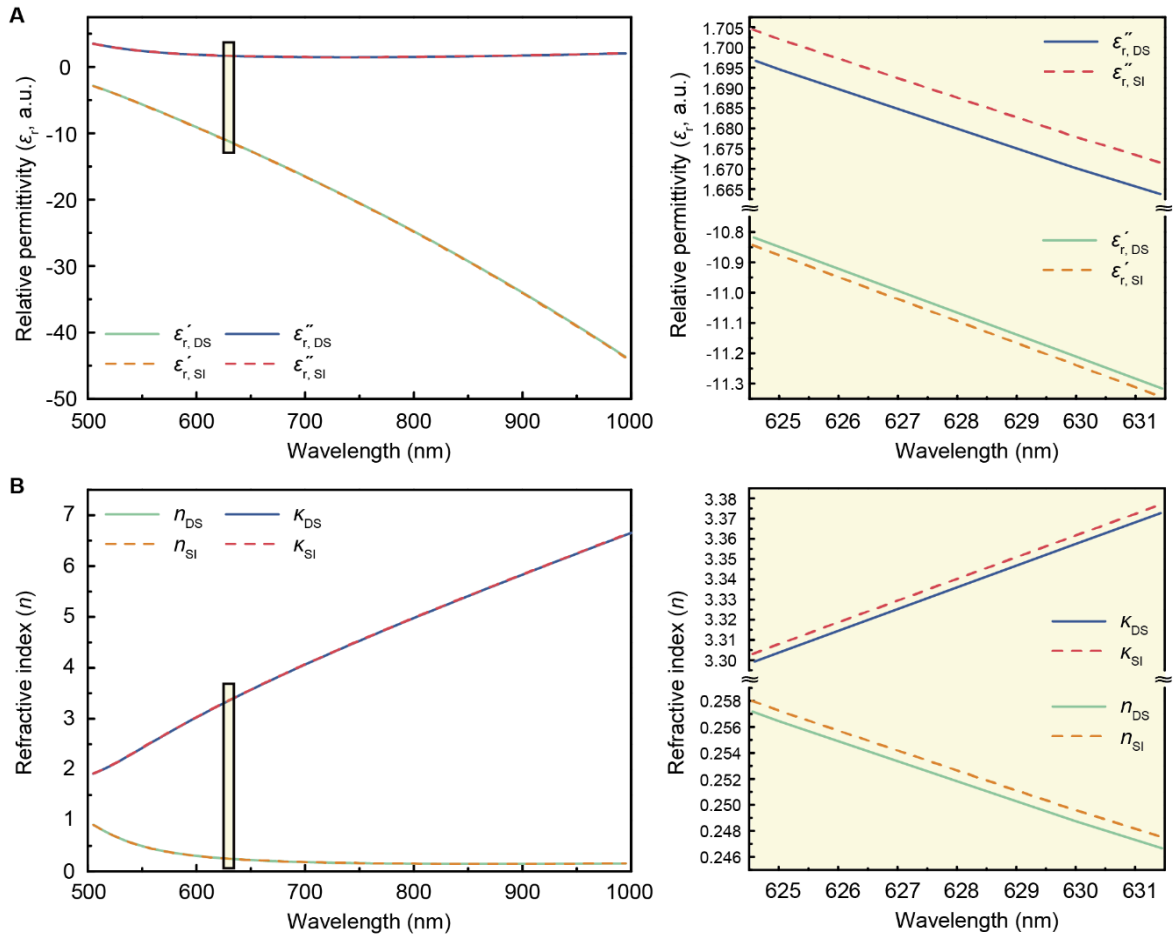

**Fig. S9. Comparison of gold dispersion between the typical Drude-Sommerfeld and strain-induced modified model (A)** comparison of dielectric function **(B)** comparison of complex refractive index.

**Fig. S10.**

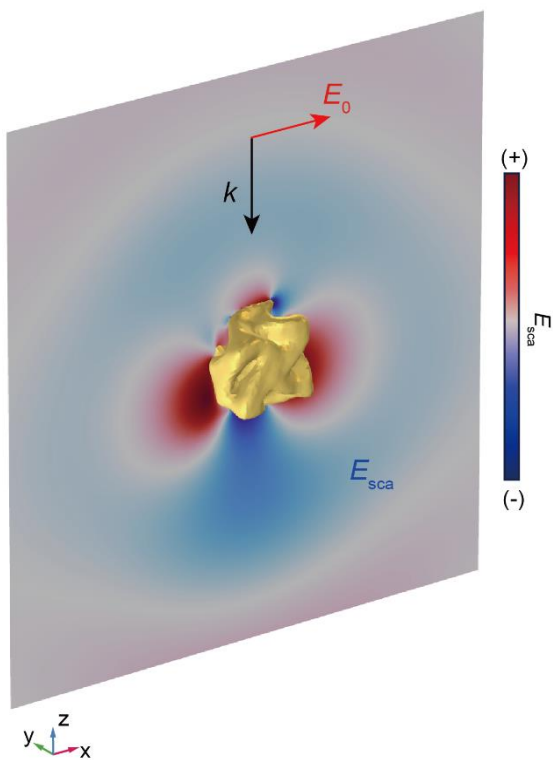
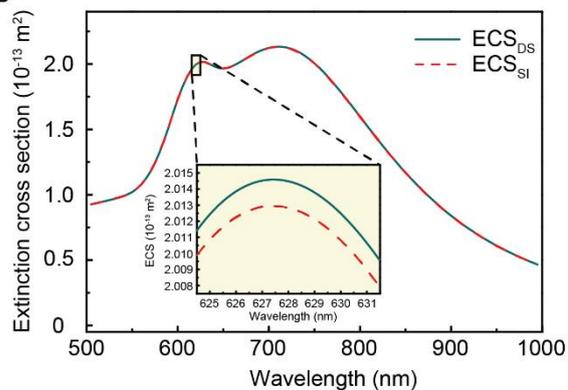
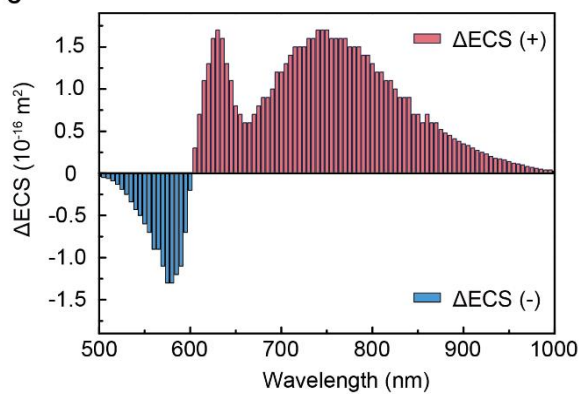

**Fig. S10. Simulated far-field extinction cross-section of 432 helicoid III nanoparticle based on BCDI result.** (**A**) The scattered electric-field of 432 helicoid III under the incidence of x-axis linearly polarized light at 630 nm (**B**) Comparison of extinction cross-section of 432 helicoid III between the Drude-Sommerfeld ($ECS_{DS}$) and strain-induced model ($ECS_{SI}$) (**C**) and the difference between two models ($ECS_{DS} - ECS_{SI} = \Delta ECS$).

**Fig. S11.**

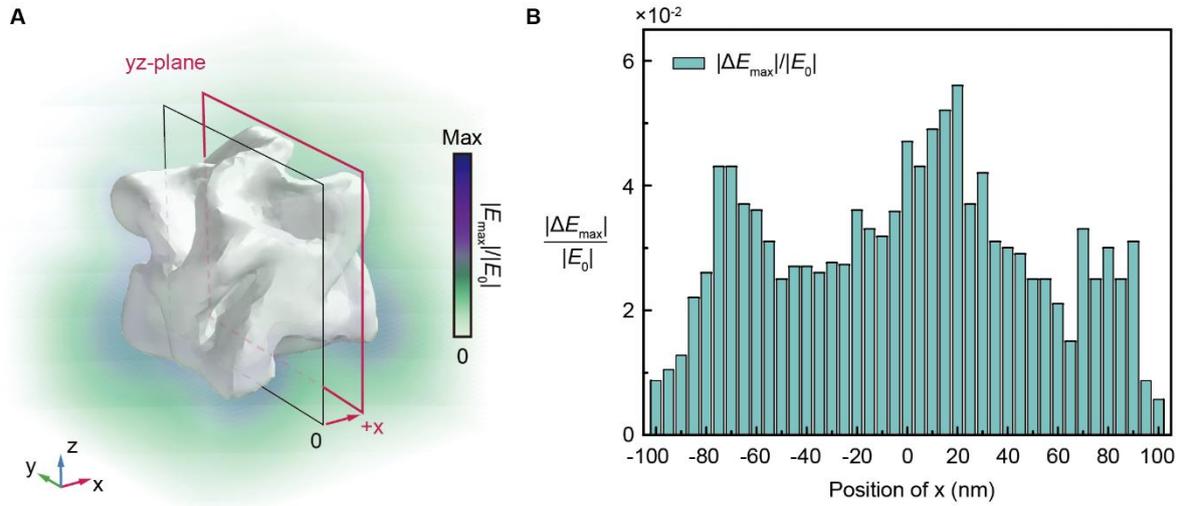

**Fig. S11. Simulated near-field distribution of 432 helicoid III nanoparticle based on BCDI result.** (**A**) The scattered electric-field distribution normalized by incidence electric-field (the incidence of x-axis linearly polarized light at 630 nm) (**B**) the difference of maximum scattered electric-fields of 432 helicoid III between the Drude-Sommerfeld ($E_{DS}$) and strain-induced model ($E_{SI}$) as the function of position x ($E_{max, DS} - E_{max, SI} = \Delta E_{max}$).

**Fig. S12.**

A 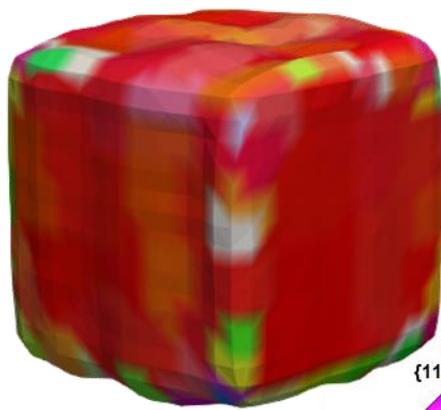 B 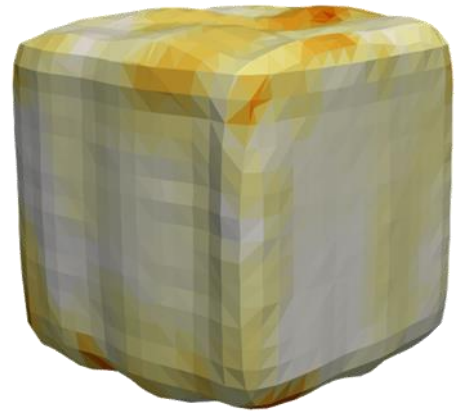

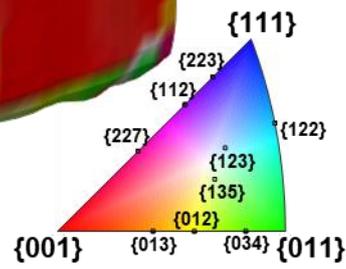

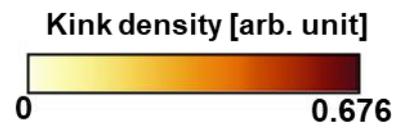

**Fig. S12. Crystallographic identification for cube nanoparticle.** (**A**) Surface Miller-index of cube gold nanoparticle. (**B**) Surface kink density of cube nanoparticle.

**Fig. S13**

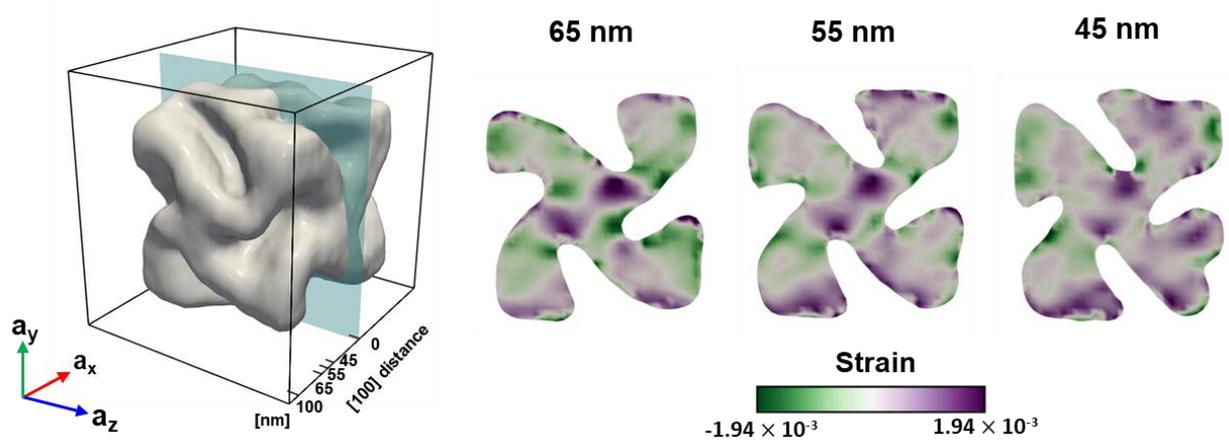

**Fig. S13. Strain analysis of other 432 helicoid III nanoparticle.** Strain slices of the typical chiral gap along with distance ($a_x$) from the center. The same symmetric strain distribution near the chiral gap is observed.

**Table S1.**

| Parameters | Value |
|---|---|
| $K$ | 0.13 |
| $v_F$ | $1.41 \times 10^6$ m/s |
| $1/\gamma_0$ | $9.3 \times 10^{-15}$ s |
| $m_{eff}$ | $0.99 \times m_{electron}$ |

**Table S1. Parameters for the dielectric constant of the nanosized bulk gold**

**Movie S1.**

**Movie. S1. Terrace area fitting procedure for cube nanoparticle.** Surface Miller-index of cube nanoparticle with varying rotation angle along $+\mathbf{y'}$ (in Fig. 1B) from -60° ~ 60° with 2° step size. At the center, the overall Miller-index of the plane becomes {100} series except the edge and corner of the nanoparticle that produces the "most red" color-coding result.